\documentclass[preprint,12pt]{elsarticle}

\usepackage{amssymb}
\usepackage{booktabs}
\usepackage{amsmath}
\usepackage{caption}
\usepackage{hyperref}
\usepackage{graphicx}
\usepackage[utf8]{inputenc}
\DeclareUnicodeCharacter{202A}{} 
\journal{Lancet}

\begin{document}

\begin{frontmatter}



\title{U-Net-based Models for Skin Lesion Segmentation: More Attention and Augmentation}


\author[inst1]{Pooya Mohammadi Kazaj}

\affiliation[inst1]{organization={K N Toosi University of Technology},%Department and Organization
            city={Tehran},
            state={Tehran},
            country={Iran}}

\author[inst2]{MohammadHossein Koosheshi}
\affiliation[inst2]{organization={University of Tehran},%Department and Organization 
            city={Tehran},
            state={Tehran},
            country={Iran}}
\author[inst3]{Ali Shahedi}
\affiliation[inst3]{organization={Amirkabir University of Technology},%Department and Organization
            city={Tehran}, 
            state={Tehran},
            country={Iran}}
\author[inst4]{‪Alireza Vafaei Sadr}
\affiliation[inst4]{organization={Uniklinik RWTH Aachen },%Department and Organization
            city={Aachen},
            state={North Rhine-Westphalia},
            country={Germany}}

\begin{abstract}
According to WHO\cite{WHO}, since the 1970s, diagnosis of melanoma skin cancer has been more frequent. However, if detected early, the 5-year survival rate for melanoma can increase to 99 percent. In this regard, skin lesion segmentation can be pivotal in monitoring and treatment planning. In this work, ten models and four augmentation configurations are trained on the ISIC 2016 dataset. The performance and overfitting are compared utilizing five metrics. 
Our results show that the U-Net-Resnet50 and the R2U-Net have the highest metrics value, along with two data augmentation scenarios. We also investigate CBAM and AG blocks in the U-Net architecture, which enhances segmentation performance at a meager computational cost. In addition, we propose using pyramid, AG, and CBAM blocks in a sequence, which significantly surpasses the results of using the two individually. Finally, our experiments show that models that have exploited attention modules successfully overcome common skin lesion segmentation problems. Lastly, in the spirit of reproducible research, we implement models and codes publicly available.\footnote{\url{https://github.com/pooya-mohammadi/unet-skin-cancer}}
\end{abstract}


%%\item Various segmentation models trained on a skin cancer dataset
%%\item Different attention modules were studied in this paper

\begin{keyword}
Deep learning \sep Skin cancer \sep Segmentation \sep U-Net \sep Attention modules \sep Data Augmentation
\end{keyword}
\end{frontmatter}
\section{Introduction}
\label{sec:intro}

Skin cancer is one of the most common diseases that must be diagnosed early. 1 in 5 Americans will develop skin cancer by the age of 70, and more than two people die of skin cancer in the United States every hour. Besides, it is a truism that an early cancer diagnosis can significantly improve a patient's chances of being cured. In this respect, Deep learning has been proven to be a pioneer in fast and accurate pattern recognition in medical applications. Consequently, it can provide specialists with valuable help during the diagnosis stages. The International Skin Imaging Collaboration (ISIC) provides high-quality dermoscopic images for classification and segmentation tasks in computer vision via deep learning. The ISIC 2016 dataset\cite{ISIC2016} is used in this work.

This paper compares models that help improve skin lesion segmentation. Models based on the U-Net architecture, like R2U-Net, Double U-Net, and U-Net Res50 are extensively explored to improve some issues that the basic U-Net model was unable to solve. Moreover, attention blocks are extensively studied. We propose U-Net CBAM Attention Gate and U-Net Pyramid CBAM Attention Gate, which improve the previous results of U-Net-based architectures. Since augmentation is an inseparable part of any deep learning research, various augmentation methods have also been investigated, such as the ones provided by the Albumentation library\cite{info11020125}. In addition, Mosaic\cite{hao2020improved} and CutMix\cite{yun2019cutmix} are tested as numerous works suggested them for reaching better-generalized models. Furthermore, two custom hair augmentation approaches are also implemented since the skin cancer images are adversely affected by human hairs located around or on the skin lesion, confusing segmentation models, and eventually leading to low-accuracy models.

In the literature on segmentation models, Dice, IoU, and Focal Tversky\cite{abraham2019novel} are widely used for evaluating the performance of the models. Therefore, in this study, we employ the difference between the training and validation metrics, $\Delta M $,  to evaluate the degree of susceptibility of a model toward overfitting. It defines to what degree a model is overfitted over training data during the training procedure. Lastly, this paper is organized as follows: in section 2, the related works of skin lesion segmentation are described. Section 3 explains the methods used in this work, namely the dataset, models, and augmentation strategies. In section 4, the evaluation methods are discussed. In section 5, the results are discussed, and models and augmentation configurations are compared. Finally, the contribution of this paper and future works are presented in section 6.

\section{Related Works}

Before the advent of deep learning, some basic machine learning algorithms were used for image classification and segmentation, such as SVM\cite{alquran2017melanoma} and Random Forest\cite{murugan2019detection}. In 2015, U-Net\cite{ronneberger2015u} was introduced and evaluated in a medical image segmentation task and proved highly promising. Since 2015, when vanilla U-Net was introduced, U-Net and U-Net-based architectures have been widely used for segmentation tasks (e.g., skin lesion segmentation). Some more robust models have been introduced recently and tested on medical segmentation challenges like ISIC\cite{ISIC}. Among these, Mask R-CNN\cite{he2017mask} and DeepLab\cite{chen2017deeplab} are to be mentioned. Bagheri et al. 2018\cite{bagheri2020two} deployed Mask-RCNN and DeepLab methods on the ISBI 2017 dataset and attained a Jaccard value of 79\%. In 2018, Zahangir Alom et al.\cite{alom2018recurrent} exploited Recurrent Residual Convolutional Neural Networks (RRCNN) on U-Net-based models and proposed R2U-Net. R2U-Net achieved a Dice Coefficient of 0.86 on the ISIC 2017 dataset. Zafar et al. 2020\cite{zafar2020skin} made use of the ResNet50\cite{he2016deep} architecture as the backbone of U-Net to reach a Jaccard index of 77\% on the ISIC 2017 dataset. Jha et al. 2020\cite{jha2020doubleu} introduced Double-U-Net combining two U-Net stacks on top of each other. The Double U-Net model gained a Dice Score of 0.8962 on the ISIC 2017 dataset. An innovation that research has recently significantly benefited from is attention modules since they provide concentration on the most important features. Kaul et al. 2019 \cite{kaul2019focusnet} presented FocusNet, an attention-based CNN proposed especially for medical segmentation. It, FocusNet, achieved a Dice Score of 0.83 on the ISIC 2017 dataset. Gu et al. 2020\cite{gu2020net} used comprehensive attention-based convolutional neural networks (CA-Net) in medical tasks to reach better accuracy. CA-Net improved the Dice Coefficient from 87\% to 92\% on the ISIC 2018 dataset.

Data augmentation was also an essential part of such studies. Perez et al. 2018\cite{perez2018data} evaluated 13 scenarios on the ISIC 2017 dataset for skin cancer classification, each of which was a combination of one or many augmentation strategies. The authors concluded that the best results come from geometry and color augmentations. The scenario with the best results was a scenario in which random cropping, rotating, random flipping, and a mixture of saturation, contrast, brightness, and hue were used, respectively. Moreover, Cutout, Cutmix, and Mixup have also been used on the ISIC 2017 dataset by French et al. 2019\cite{french2019semi}. Their experience declared that Cutmix could set the path to better segmentation results than the other two.

\section{Methods}
This section describes all technical details about the implementation of models. It containes three subsections, Dataset, Models, and Augmentations. We introduce the dataset and present the proposed models reviewed in the dataset and model sections. Finally, in the Augmentation section, the four augmentation scenarios proposed by the authors are elaborated. 
\subsection{Dataset}
\begin{figure}[t]
    \centering
    \includegraphics[width=75mm, scale=0.5]{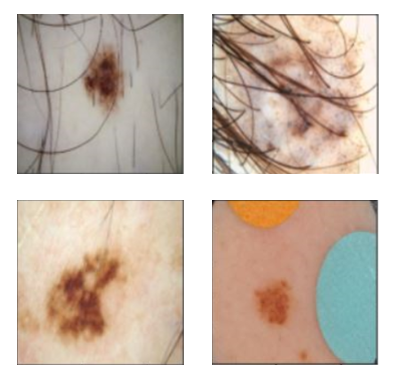}
    \caption{four samples from the ISIC 2016 dataset\cite{ISIC2016}}
    \label{fig:ISIC_samples}
\end{figure}

In this project, the models are trained on the ISIC 2016 Dataset\cite{ISIC2016}. The training dataset contains 900 dermoscopic images, each containing a skin lesion in JPEG format and related mask images in PNG format. The testing dataset contains 379 images of the same format as the training set(for both images and related masks). Firstly, training data is split into 720 and 180 images as training and validation sets, respectively. In this dataset, images have various shapes for both height and width, so they are resized to 256 by 256. In figure~\ref{fig:ISIC_samples} shows four samples of the ISIC 2016 dataset. The first row shows skin lesions images that are covered by hairs. These hairs adversely affect the model's performance and are highly challenging for segmentation models. Two strategies are explained in the augmentation section to address these artifacts.

\subsection{Models}
In this section, the implemented models are introduced. U-Net models are one of the most famous structures in image segmentation tasks and have been widely used on the ISIC 2016 dataset. Accordingly, this work is directed to evaluate the most effective U-Net models, especially those that were state-of-the-art for image segmentation tasks. Furthermore, attention modules are also tested to enhance the existing models' performance on the ISIC 2016 dataset. In the following, three models that produce the best results, along with the model that the authors propose, are explained. All the models are summarized in table~\ref{table:models}. Note that the model sizes are equal to the sizes of each model's weight after being saved on a hard drive in .h5 format, which is the default format for saving the Keras/TensorFlow-based\cite{TensorFlow} models. 

%% add reference for models
\begin{table}
\centering
\caption{Information of the models examined in this study. }
\captionsetup{justification=centering}
\label{table:models}
\begin{tabular}{|llll|}
\toprule
Label &            Model name  & \# parameters[M] & \# Model size[MB] \\
\midrule
R2UC &             r2unet cbam  &    25.4     &   102.1      \\
R2U  &                  r2unet\cite{alom2018recurrent}  &    96.1     &   384.7      \\
UR50 &              unet res50  &    20.7     &   83.1       \\
U-Net  &        unet conv deconv\cite{ronneberger2015u}  &    7.7      &   31.0       \\
UAG  &     unet attention gate\cite{oktay2018attention}  &    0.8      &   3.6        \\
UC   &               unet cbam  &    7.7      &   31.2       \\
UCG  &          unet cbam gate  &    0.9      &   3.7        \\
UPCG &  unet pyramid cbam gate  &    4.4      &   17.8       \\
MCGU &                mcg unet\cite{asadi2020multi}  &    1.7      &   6.9        \\
DU   &             double unet\cite{jha2020doubleu}  &    29.3     &   117.7      \\
\bottomrule
\end{tabular}
\end{table}

\subsubsection{U-Net-Resnet50}
\begin{figure}[ht]
\centering
\includegraphics[width=50mm,scale=0.5]{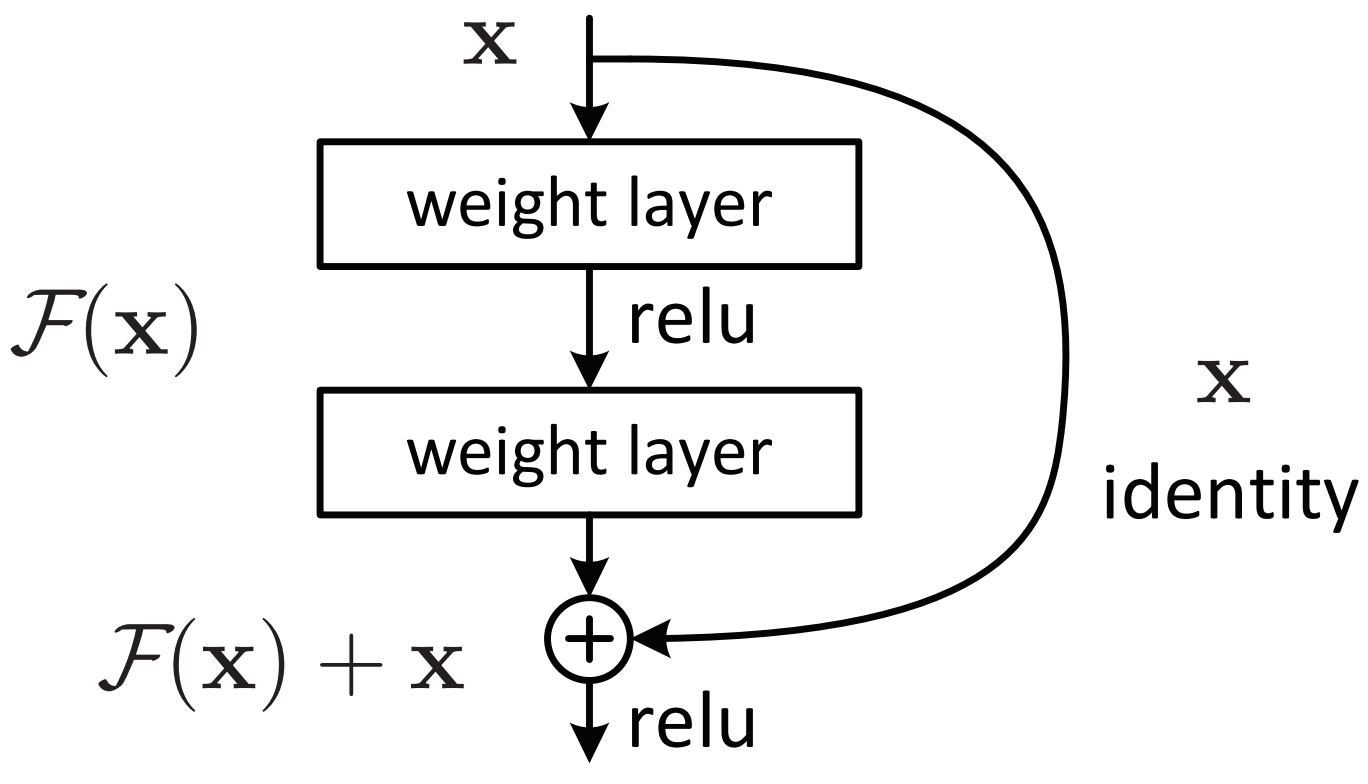}
\caption{A residual block used in ResNet50 architecture\cite{he2016deep}.}
\label{fig:residual}
\end{figure}

The first model that is proposed in the Models section is U-Net-Resnet50. Primarily, Resnet50\cite{he2016deep} was proposed to overcome problems with training deep neural networks with numerous layers, which are prone to computational complexity and more errors. A schematic figure of a Residual function is shown in figure~\ref{fig:residual}. As it's evident in the figure, adding the identity to the output helps to reduce errors and ease the training process. The ResNet50 consists of 50 convolution layers that are improved with the above-mentioned Residual blocks. The U-Net model\cite{ronneberger2015u} consists of two encoder and decoder units. The encoder unit can be built from scratch (such as the U-Net Conv Deconv model mentioned in table~\ref{table:models}). However, it is usually replaced with a successful pre-trained classification network. The U-Net-Resnet50 is mainly based on U-Net, except it uses Resnet50 as its backbone. 

\subsubsection{R2U-Net}

\begin{figure}[ht]
\centering
\includegraphics[width=100mm,scale=1]{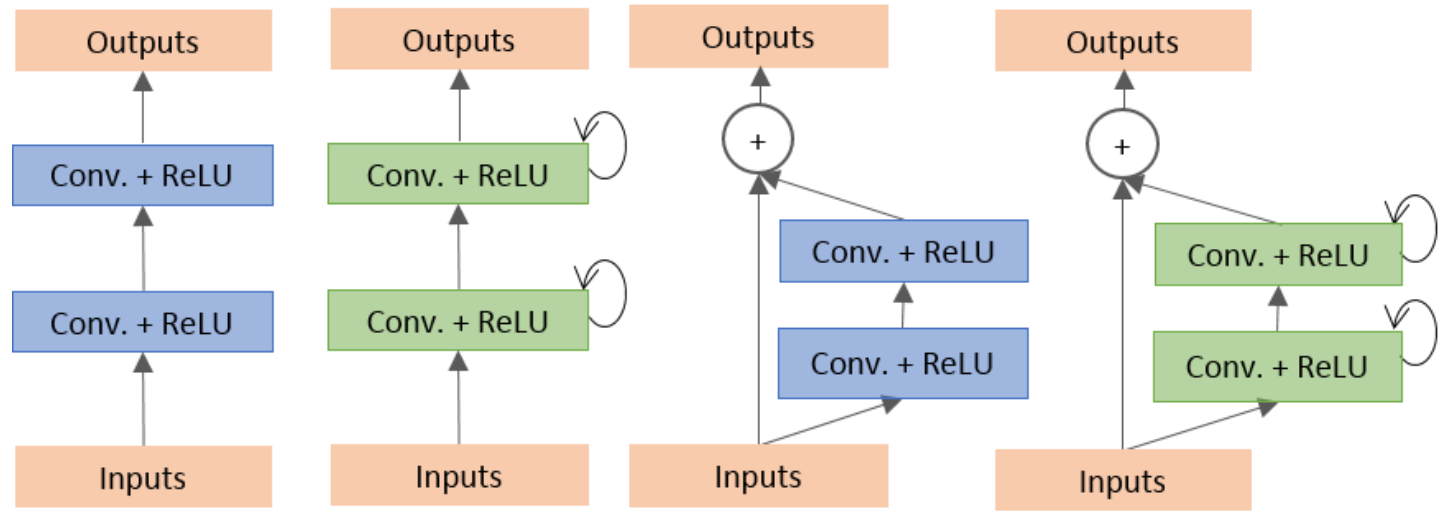}
\caption{The left image shows a recurrent network, the image in the middle depicts a residual network, and the one on the right shows a recurrent residual network\cite{alom2018recurrent}.}

\label{fig:r2unet}
\end{figure}

Recurrent Residual U-Net (R2U-Net) is a combination of residual neural networks (ResNet) and recurrent neural networks (RNN). Residual networks employ several skip connections or shortcuts as a way to skip an intermediary block. These networks simplify the training process by reducing the number of network layers; therefore, the network learns faster during the initial stages. Many popular deep learning networks have emerged from such ideas, like the ResNet50 model that was explained previously. Recurrent neural networks (RNN)\cite{lipton2015critical} are another example of renowned deep learning structures. Artificial neurons in this deep network are adjusted by their memory of their previous inputs. In other words, the inputs and outputs depend on each other since the model utilizes the previous inputs as its memory to predict the following inputs. Given these two methods, R2U-Net\cite{alom2018recurrent} has been introduced to utilize both mentioned networks' advantages: a faster learning process and better pattern recognition. R2U-Net is mainly proposed to overcome problems related to medical image segmentation, as in this work. A depiction of the three mentioned structures is provided in figure~\ref{fig:r2unet}. 

\subsection{U-Net CBAM}

\begin{figure}[ht]
\centering
\includegraphics[width=100mm,scale=1]{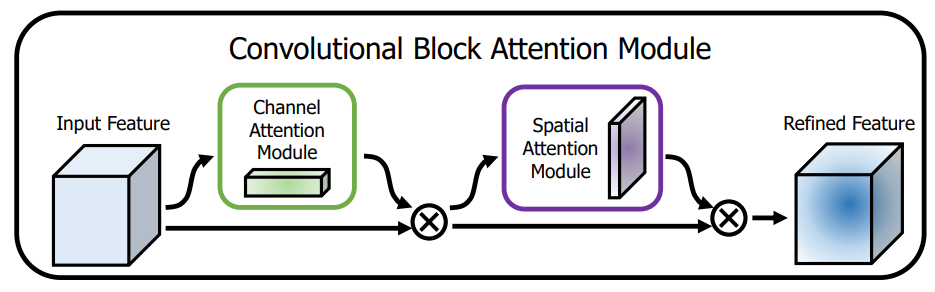}
\caption{The mechanism of the CBAM block including the channel and spatial modules\cite{woo2018cbam}}
\label{fig:cbam}
\end{figure}

In recent research studies, attention modules are used to help CNNs to focus on more significant features from input images and not get distracted by less important ones. Standing for Convolutional Block Attention Module, CBAM\cite{woo2018cbam} consists of two sequential sub-models: channel and spatial. The channel module is a 1D attention map, and the spatial module is a 2D attention map. The output of the channel module, the first block, also called intermediate output, is an element-wise multiplication of the input features and the output of the channel module. This intermediate output is fed to the spatial module. The final refined output of the CBAM block is an element-wise multiplication of the intermediate output and the output of the spatial module. The mechanism of the CBAM block can better be understood through figure~\ref{fig:cbam}.

Putting the CBAM block in the way of skip connections of U-Net is also studied in this paper. According to our early experiments, U-Net cannot make better segmentation in more challenging areas, such as borders, darker skin tones, more complex skin lesion patterns, etc. By putting CBAM in skip connection layers, features in the encoder unit are transformed into a feature map of richer features, leading to better segmentation performance.

\subsection{U-Net CBAM Attention Gate}

\begin{figure}[ht]
\centering
\includegraphics[width=150mm,scale=1]{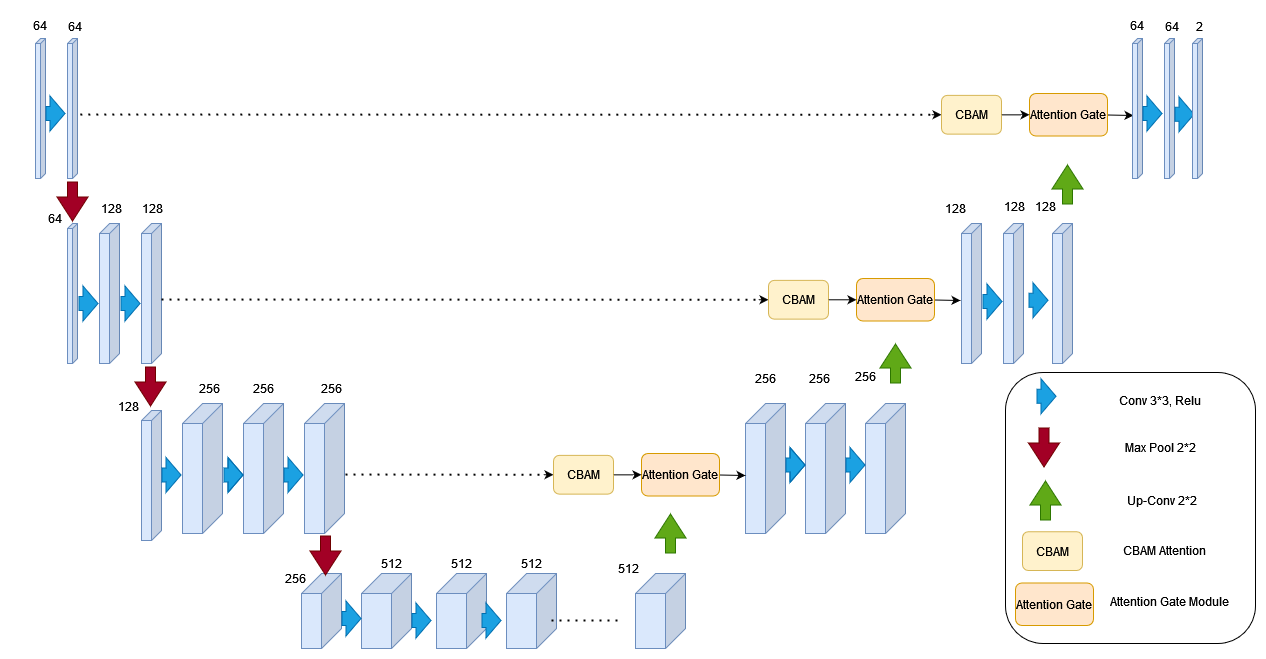}
\caption{The architecture of U-Net CBAM Attention Gate model}
\label{fig:unetcbamgate}
\end{figure}

Attention modules can help U-Net architectures to learn better by focusing on more important features. Proposed by \cite{oktay2018attention}, attention gate (AG) scales the input features with attention coefficients, which are computed through additive attention, inspired by\cite{bahdanau2014neural}. In other words, areas of higher significance are weighted more than the less significant ones. Consequently, the areas with higher weights get more attention during training. In the U-Net architecture, the skip connections provide the upsampling section with spatial information from the downsampling section. One can mitigate this by using attention gates along the skip connection layer. In the U-Net architecture, an attention gate gets two inputs. One from the previous lower layer of the upsampling/decoder section of the network as its coarser gating signal, and the other from the downsampling/encoder section through a skip connection layer. We propose using the CBAM block and the AG in a sequence. The CBAM block can help the Attention U-Net to make use of higher-level features than before. As a result, the U-Net architecture performs segmentation more robustly, benefiting from both modules' advantages. An illustration of the UCG model is provided in figure~\ref{fig:unetcbamgate}

\subsubsection{R2U-Net-CBAM}

This model makes use of CBAM in the way of skip connections of R2U-Net. Our experiments show that R2U-Net lacks accuracy in some critical situations, e.g., edges of a skin lesion, and that attention modules can be of great help. Furthermore, CBAM is an effective and efficient attention block. To the authors' best knowledge, R2U-Net-CBAM has not been suggested before.

\subsubsection{U-Net Pyramid CBAM Attention Gate}

\begin{figure}[ht]
\centering
\includegraphics[width=150mm,scale=1]{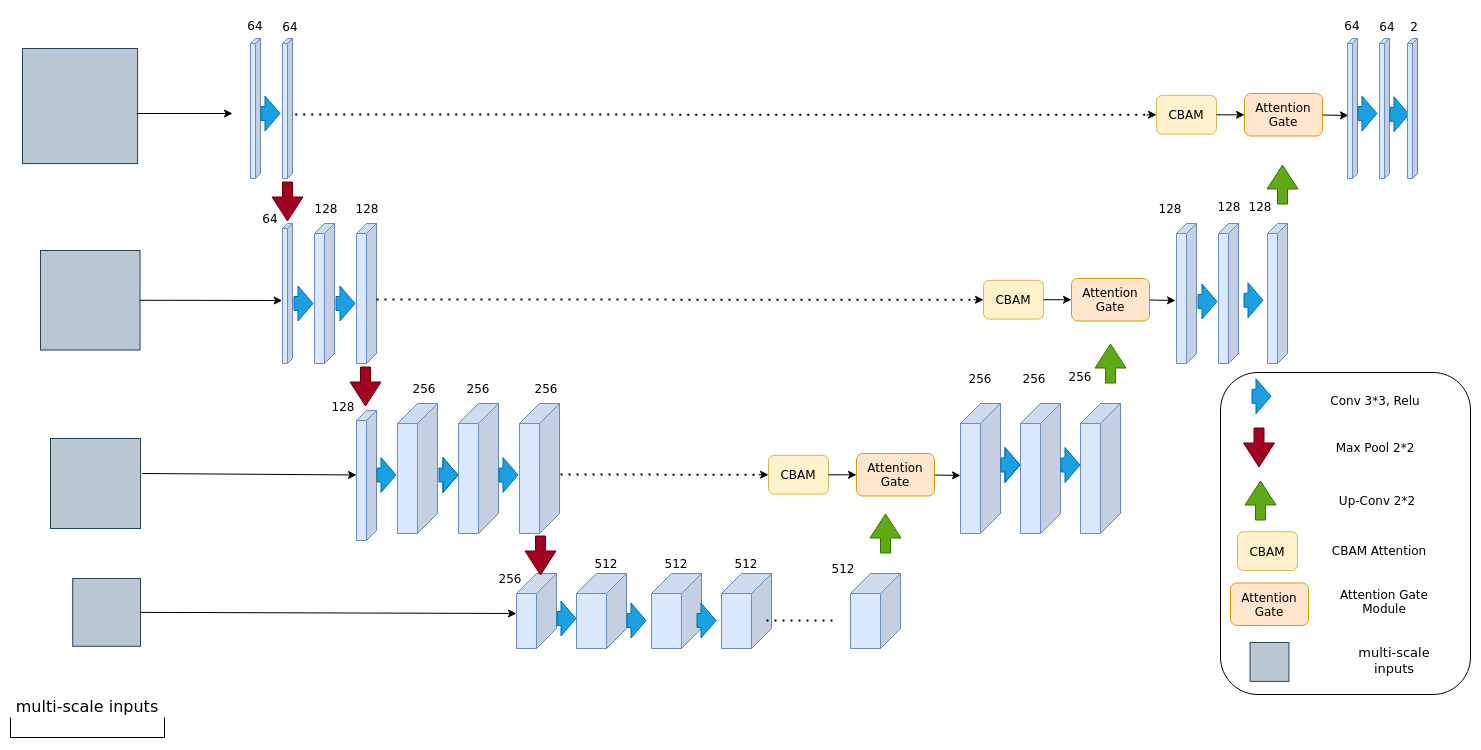}
\caption{The architecture of U-Net Pyramid CBAM Attention Gate model. Feature pyramids are provided to the U-Net model to improve resolution in the network}
\label{fig:UPCG}
\end{figure}

As convolutional neural networks get deeper, although feature representation increases, the resolution decreases. Pyramid inputs have been offered to overcome this problem. The idea is to feed the multi-scaled inputs, which are constructed by average pooling blocks, to the different levels of the encoding section of the U-Net architecture. Input image pyramid and attention gates in the U-Net model were used by \cite{abraham2019novel} on skin lesion images. In our U-Net Pyramid CBAM Attention Gate model, the sequence of CBAM blocks and attention gates are used along with multi-scale input images. The model is illustrated in figure~\ref{fig:UPCG}.

\subsection{Augmentations}
Reaching generalized models has always been a challenging task. Accordingly, in this study, some parameters are set to augment the input images from the dataset before entering the models to increase the generalization of the investigated models. Corresponding probabilities can modify these parameters. Based on \cite{perez2018data}, a combination of random crops, rotation, flips, saturation, contrast, brightness, and hue can generate satisfactory results. Therefore, this work uses rotation and flips along with Cutmix and Mosaic as alternatives to random crops because they could promise better generalization in skin lesion segmentation tasks.

\begin{figure}[ht]
\centering
\includegraphics[width=100mm,scale=0.8]{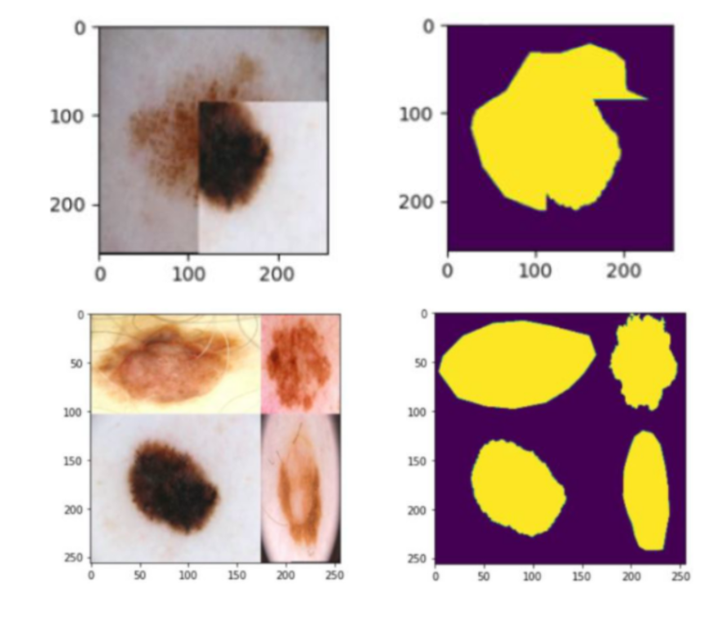}
\caption{Top row: Cutmix augmentation applied on an image from the ISIC 2016 dataset, Bottom row: Mosaic augmentation applied on an image from the ISIC 2016 dataset}
\label{fig:augment}
\end{figure}

Cutmix cuts a random patch of an image and pastes it on a random patch of another random image. The same patches are also cut and pasted from the ground truth mask images. In addition, mosaic data augmentation combines four training images and related ground truth masks into one image and a ground truth mask in random ratios. Moreover, hair augmentation and removal methods are proposed to address the problems with the appearance of hair in the training images. Implementation of Cutmix and Mosaic methods are depicted in \ref{fig:augment}

\begin{figure}[ht]
\centering
\includegraphics[width=75mm,scale=0.8]{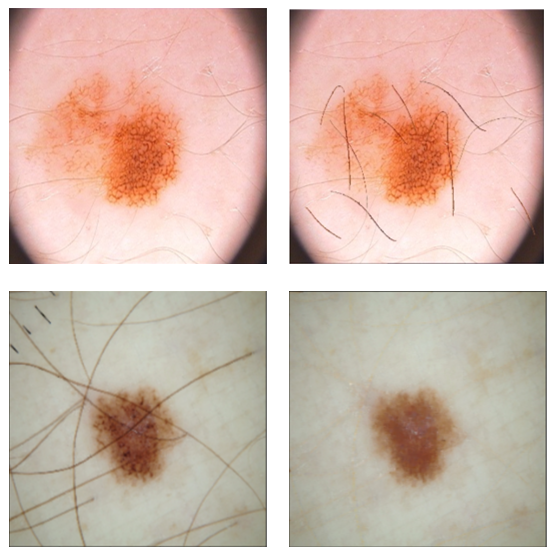}
\caption{Top Left: a standard image from the ISIC 2016 dataset, Top Right: the image with added artificial hairs as hair augmentation, Bottom Left: a standard image from the ISIC 2016 dataset, Bottom Right: the image after removing its hairs}
\label{fig:hair}
\end{figure}

Many images in the ISIC datasets are affected by body hairs, making it hard for models to discriminate lesions from hairs since they have the same appearance. Thus, segmentation models will be biased toward these hairs and may label them as lesions or even leave out some lesions, thinking they are side effects of body hairs. In hair augmentation, artificial hairs with the same appearance as body hairs are applied to the input images and not to the masks. As a result, models lose their bias toward hairs because they see them in many entry images, not in the corresponding mask images. Hair removal is another approach to resolving models' sensitivity toward hairs on skin lesions. This augmentation method removes the hair objects from the images. Figure~\ref shows the implementation of hair removal and augmentation is shown in figure~\ref{fig:hair}. For convenience, all four augmentation scenarios investigated in this study are presented in table~\ref{table:aug}.

\begin{table}
\centering
\caption{four different augmentation configurations used in this study}
\label{table:aug}
\begin{tabular}{|lllllll|}
\toprule
 Label & Rotation & H/V Flip & CutMix & Mosaic & Hair-Aug & Hair-removal \\
\midrule
 AUG-1 &  180 & 0.5      &  No & No &   No &  No          \\
 AUG-2 &  180 & 0.5      &  Yes & Yes &  No &  No          \\
 AUG-3 &  180 & 0.5      &  Yes & Yes &  Yes &  No          \\
 AUG-4 &  180 & 0.5      &  Yes & Yes &  No &  Yes          \\
\bottomrule
\end{tabular}
\end{table}

\section{Evaluation}
In this section, the evaluation process and the used metrics are explained. Each model is trained on 900 training images and then evaluated on 379 test images, which the competition organizers split. Each set is augmented with one of the four types of the proposed augmentation sets. This procedure is repeated five times for each setup to reduce the uncertainty of the results. Then, they are averaged to derive the mean and the standard variation of each experiment containing a model and an augmentation strategy discussed earlier. For the evaluation of the models and augmentations, three commonly used metrics for segmentation tasks are used. These metrics are Dice Score, Focal Tversky, and IoU, which are computed as follows:

\begin{equation}
\centering
IoU(A,\,B) = \frac{A\cap B}{A\cup B} = \frac{\text{area of overlap}}{\text{area of union}}
\end{equation}
where A, and B denote ground truth area and prediction area, respectively.

\begin{equation}
\centering
Dice Score (y,\,\hat{p})=\frac{2 y \hat{p}+1}{y+\hat{p}+1}
\end{equation}
where y, $\hat{p}$ denote ground truth area and prediction area, respectively.

\begin{equation}
\centering
  \label{eq:t}
  \begin{aligned}
    T L(p, \hat{p})=1-\frac{1+p \hat{p}}{1+p \hat{p}+\beta(1-p) \hat{p}+(1-\beta) p(1-\hat{p})}, \\   
    F T L=\sum_{c}\left(1-TL_{c}\right)^{\gamma}
  \end{aligned}
\end{equation}

Where FTL, TL, p, $\hat{p}$, $\beta$, and $\gamma$ denote Focal Tversky Loss, Tversky Loss, ground truth area, prediction area, beta value, and gamma value, respectively. This work sets gamma and beta to 0.75 and 0.7, respectively.  
The difference between training and validation results is used to measure model generalization called $\Delta M $. Since this metric computes the exact gap between validation and training metrics, it is a simple but intuitive way to demonstrate a model's generality. The lower a model can keep this metric, the higher its generality. This metric is computed with the following equation:

\begin{equation}
\centering
    \Delta M = \lvert M_{Trn}-M_{Val} \rvert
\end{equation}
where $M$, $M_{Trn}$, and $M_{Val}$ denote the metric that could be Dice, FT, or IOU, the training metric, and the validation metric, respectively.

Finally, A metric is proposed to monitor models' training speed. The training speed is one of the main characteristics of a model, which is hard to measure. A metric threshold is defined to evaluate this metric quantitatively. The first epoch that a model crosses this threshold is reported as its training speed. The lower the epoch number in which a model crosses the predefined threshold, the higher its training speed. In this work, the threshold of the Dice Score for measuring the training speed is set to 0.8. Therefore, if a model's Dice Score crosses 0.8 at epoch x, x is reported as its speed identifier.

\section{RESULTS}
\label{sec:res}

This section provides the results and comparison of ten studied models and four distinct augmentations. For training each model, a batch size of 32 is selected, and each setup is trained for 60 epochs, while the early stopping is set to stop the training if there is no improvement for 30 consecutive epochs. Additionally, stochastic gradient descent with 0.9 momentum is chosen as the optimizer, while the initial learning rate is set to 0.01. If there is no improvement for 15 successive epochs, the learning rate is decreased by 0.1. Furthermore, Dice Loss is used as the loss function of all the setups. Finally, the training is performed on 720 training images and evaluated on 180 validation images to segment skin lesions in the ISIC 2016 dataset.

\begin{table}
\centering
\caption{Model results averaged over four augmentation configurations}
\label{table:all-models-agus}
\begin{tabular}{|l|ll|ll|ll|}
\toprule
{} &       dice &   val\_dice &         FT &     val\_FT &        iou &    val\_iou \\
\midrule
R2UC &   0.91±0.0 &  0.88±0.01 &  0.17±0.01 &  0.21±0.02 &  0.83±0.01 &  0.79±0.02 \\
R2U  &   0.91±0.0 &  0.88±0.01 &   0.17±0.0 &   0.2±0.01 &   0.83±0.0 &  0.79±0.01 \\
UR50 &   0.91±0.0 &   0.91±0.0 &   0.16±0.0 &  0.17±0.01 &   0.84±0.0 &  0.83±0.01 \\
U-Net  &  0.81±0.02 &  0.79±0.02 &  0.29±0.02 &   0.3±0.02 &  0.68±0.03 &  0.66±0.03 \\
UAG  &   0.8±0.03 &  0.79±0.04 &   0.3±0.03 &   0.3±0.04 &  0.67±0.04 &  0.66±0.05 \\
UC   &  0.82±0.02 &  0.81±0.01 &  0.28±0.02 &  0.29±0.02 &  0.71±0.03 &  0.69±0.02 \\
UCG  &   0.8±0.03 &  0.79±0.03 &   0.3±0.03 &   0.3±0.03 &  0.68±0.04 &  0.66±0.04 \\
UPCG &  0.88±0.01 &  0.86±0.01 &  0.21±0.01 &  0.22±0.02 &  0.79±0.01 &  0.76±0.02 \\
MCGU &  0.82±0.02 &  0.79±0.03 &  0.27±0.02 &  0.29±0.03 &   0.7±0.03 &  0.66±0.04 \\
DU   &   0.89±0.0 &  0.89±0.01 &  0.18±0.01 &  0.18±0.01 &  0.81±0.01 &  0.81±0.01 \\
\bottomrule
\end{tabular}
\end{table}
Table~\ref{table:all-models-agus} shows the best-obtained train and validation metrics averaged over four augmentation setups for ten various deep neural network architectures. According to the table, U-Net-Resnet50 obtains the highest validation Dice score of 0.91. At the same time, Double-U-Net, R2U-Net, and R2U-Net-CBAM present remarkable performances and obtain the second, the third, and the fourth highest Dice scores of 0.89, 0.88, and 0.88, respectively. This performance superiority of the U-Net-Resnt50 model is mainly attributed to its backbone, extracting essential and rich feature sets from the input images and passing them to the decoder section of the model. At the same time, it has the lowest model size compared to its three counterparts. In terms of comparing models that benefit from attention modules, it is notable that the CBAM block improves the results of the conventional U-Net by roughly 3\% IoU score from 68\% to 71\%. In addition, using the pyramid sequence, CBAM, and attention gate modules, the performance remarkably improves and reaches from a Dice Score of 0.82 for U-Net-CBAM to 0.88.      

Likewise, table~\ref{table:all-augmentation-models} demonstrates the best-obtained train and validation metrics of four augmentation setups averaged over all ten studied models. According to this table, Aug-1 and Aug-2 setups have the highest average dice scores of 0.87. The results show that the Hair-Augmentation and Hair-Removal are ineffective in skin lesions segmentation and may even decline in performance. The authors hypothesize that hairs' accidental removal and augmentation create more confusion at the pixel level than making the models more robust by visiting more general training images. In other words, these two augmentation methods create artificial artifacts on the training images that diverge the models' attention from the actual hairs and skin lesions they should focus on. It is because the segmentation models focus on a pixel level of the training image versus the classification models that focus on the overall image to classify it. 

\begin{table}
\centering
\caption{Augmentation results averaged over all ten models}
\label{table:all-augmentation-models}
\begin{tabular}{|l|ll|ll|ll|}
\toprule
{} &       dice &   val\_dice &         FT &     val\_FT &        iou &    val\_iou \\
\midrule
AUG-1 &  0.89±0.05 &  0.87±0.05 &  0.19±0.06 &  0.21±0.05 &  0.81±0.07 &  0.77±0.07 \\
AUG-2 &  0.88±0.04 &  0.87±0.04 &   0.2±0.05 &  0.22±0.05 &   0.8±0.06 &  0.77±0.06 \\
AUG-3 &  0.88±0.04 &  0.86±0.04 &   0.2±0.05 &  0.22±0.05 &  0.79±0.06 &  0.77±0.06 \\
AUG-4 &  0.88±0.04 &  0.86±0.04 &   0.2±0.05 &  0.22±0.05 &  0.79±0.06 &  0.76±0.06 \\
\bottomrule
\end{tabular}
\end{table}

To elaborate more on the results, table ~\ref{table:learning speed} depicts all the models' training speeds. In this table, all the model's epoch numbers in which they cross the threshold dice score of 0.8 are presented. Based on the table, R2U-Net has the highest training speed in reaching the predefined threshold at about epoch 8 with zero failures. Zero failures express that the model has reached this threshold in all five training setups. Similarly, R2U-Net-CBAM, with a difference of 1 epoch, is slightly slower than R2U-Net but shows higher stability because of its lower standard variation. Finally, the third-fastest model regarding the training speed is U-Net-Resnet50 which passes the defined threshold at around epochs 12 and 13 with the lowest standard variation compared to all the other models. 

\begin{table}
\centering
\caption{Models' training speed indicated by epoch number in which they pass a predefined threshold of 0.8 Dice Score}
\label{table:learning speed}
\begin{tabular}{|l|rrr|}
\toprule
{} &  mean-epoch &  std-epoch &  failures \\
     &             &            &           \\
\midrule
DU   &        27.8 &        3.9 &       0.0 \\
MCGU &        39.0 &        6.4 &       1.0 \\
R2U  &         7.8 &        3.1 &       0.0 \\
R2UC &         8.8 &        1.5 &       0.0 \\
UAG  &        54.5 &        3.5 &       3.0 \\
UC   &        55.0 &        2.2 &       1.0 \\
UCG  &        56.0 &        1.6 &       2.0 \\
U-Net  &        56.7 &        1.2 &       2.0 \\
UPCG &        22.8 &        7.3 &       0.0 \\
UR50 &        12.6 &        0.5 &       0.0 \\
\bottomrule
\end{tabular}
\end{table}

\begin{figure}[ht]
\centering
\includegraphics[width=\linewidth]{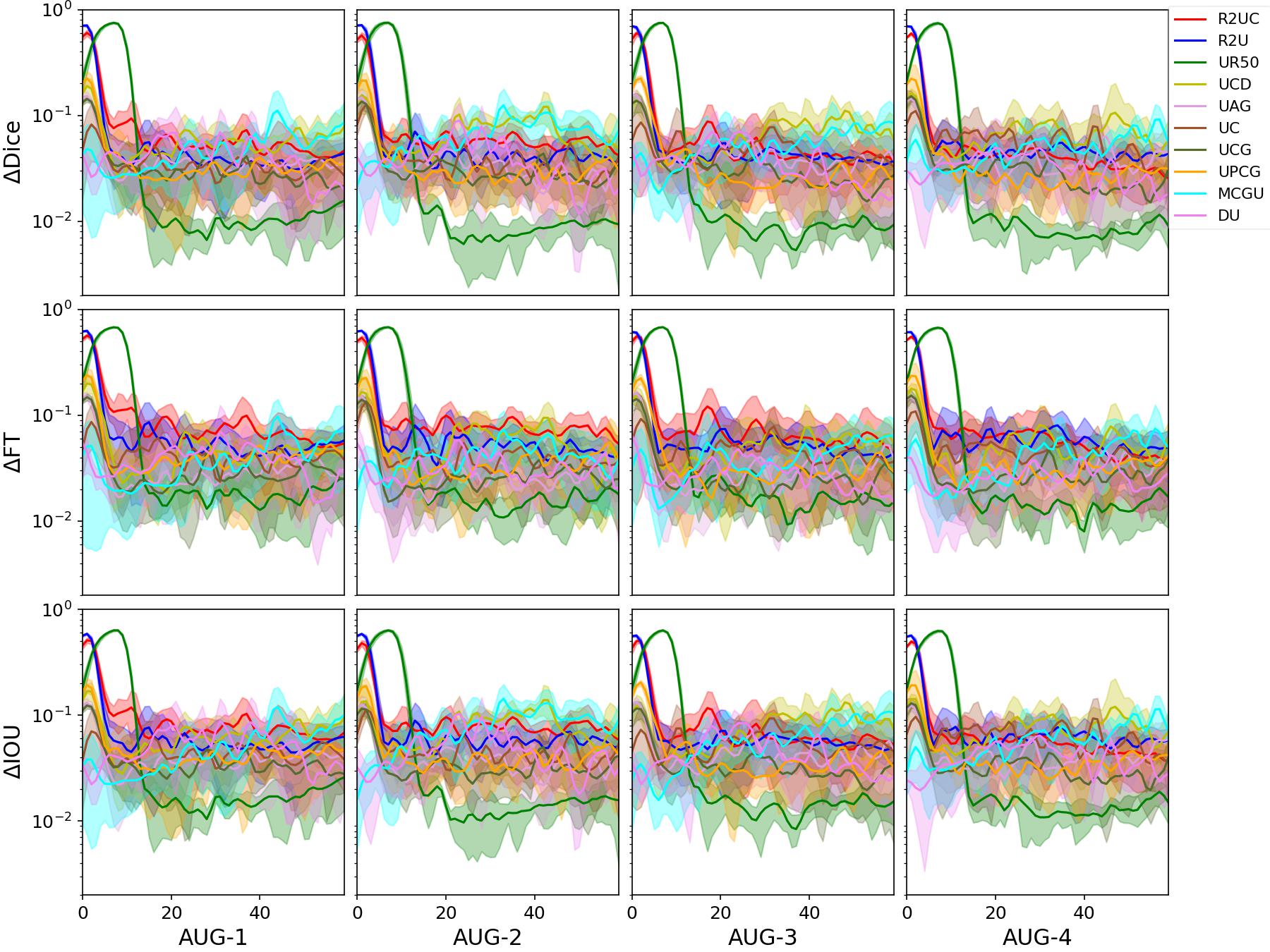}
\caption{Comparison of overfitting metrics using different loss functions for all the models and augmentation configurations throughout the training process.}
\label{fig:delta}
\end{figure}

As the final metric, the overfitting or $\Delta M $ values of all the models and augmentation setups are illustrated in figure~\ref{fig:delta} and table~\ref{table:overfitting_metrics}. Despite the high overfitting values of the U-Net-Resnet50 at the beginning of the training compared to other models, it manages to reduce its overfitting metric to the lowest value compared to other models starting at about epoch 17 with each of the four different augmentation setups. It indicates that this model is the best to prevent overfitting. 

\begin{table}
\centering
\caption{Overfitting values for all the models after epoch 15, averaged over four augmentation configurations}
\label{table:overfitting_metrics}
\begin{tabular}{|l|l|l|l|}
\toprule
{} &         Dice &           FT &          IOU \\
\midrule
R2UC &  0.048±0.007 &  0.066±0.008 &   0.066±0.01 \\
R2U  &  0.041±0.002 &  0.053±0.002 &  0.057±0.002 \\
UR50 &  0.009±0.001 &  0.016±0.001 &  0.015±0.001 \\
U-Net  &  0.067±0.006 &  0.051±0.005 &  0.077±0.007 \\
UAG  &  0.035±0.003 &  0.029±0.002 &   0.04±0.003 \\
UC   &  0.041±0.007 &  0.042±0.003 &  0.051±0.008 \\
UCG  &  0.027±0.001 &  0.026±0.002 &  0.031±0.001 \\
UPCG &  0.028±0.002 &  0.034±0.004 &  0.036±0.003 \\
MCGU &  0.061±0.008 &  0.045±0.001 &  0.072±0.011 \\
DU   &  0.038±0.004 &  0.033±0.004 &   0.05±0.005 \\
\bottomrule
\end{tabular}
\end{table}

Moreover, table~\ref{table:overall} shows the five metrics used as comprehensive model evaluations. Based on this table, the U-Net-Resnet50 has obtained the best four out of five metrics which shows its superiority over the other models. Eventually, as mentioned earlier and depicted in figure~\ref{fig:augs}, among the augmentation setups in table ~\ref{table:aug}, Aug-1 and Aug-2 get the highest scores. Therefore, mentioned models in table ~\ref{table:overall} and the two augmentations setups are the best candidates for lesion segmentation tasks.

It is essential to visualize the performance of deep segmentation models and not to delimit judgment to metrics. According to \cite{yu2018learning}, the most prevalent mistakes in segmentation are categorized into two classes: Intra-class inconsistency and inter-class indistinction. Intra-class inconsistency occurs when two patches have the same label but are labeled differently due to different appearances. In inter-class indistinction, two areas are labeled differently, but their similar appearances lead the model to label them as similar. These two problems are mainly seen in skin lesion segmentation experiments. About the intra-class inconsistency, there are parts of lesions with different appearances (mostly color and especially when lesions become lighter in color tone). However, still, they need to be labeled as a lesion while the model predicts wrongly. Moreover, about inter-class indistinction, sometimes there are other patterns on the images that are not labeled as lesions but have the same color as a lesion. Some of our other problems, such as hair appearances in the images, patients with darker skin tones, and the most significant one, and poor border detection, could also be classified as inter-class indistinction.

\begin{figure}[ht]
\centering
\includegraphics[width=\linewidth]{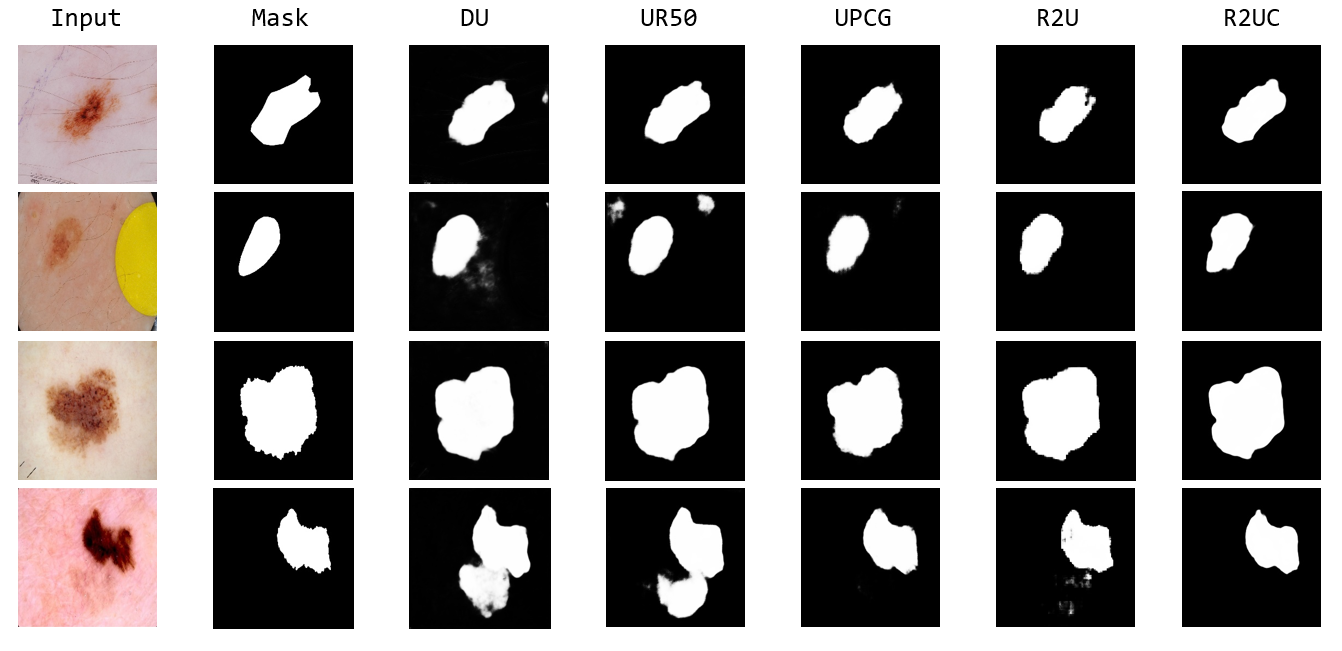}
\caption{Four experiments on the test set of the ISIC 2016 dataset using the top five models of this study. }
\label{fig:exp}
\end{figure}

In figure~\ref{fig:exp}, some tests on the test set of the ISIC 2016 dataset are presented. In the first row of figure~\ref{fig:exp}, it can be seen that the UR50 and R2UC models suggested the best segmentation results. The Double U-Net model did not perform as well in borders. In addition, it is clear that the R2U model suffers from intra-class inconsistency, but the R2UC has overcome this problem and made more robust predictions in the top right part of the lesion. The second row of figure~\ref{fig:exp} illustrates another experiment in which the UR50 model showed inter-class indistinction since it wrongly predicted some parts of the skin that had the same appearance as the skin lesion but had different labels to be skin lesions. The UPCG model performed better than the DU and UR50 models. The third row of the figure~\ref{fig:exp} shows an example in which the R2UC model performed more delicately than the R2U in the borders.
Additionally, the UPCG model outperformed the UR50 in some examples. In the fourth row of figure~\ref{fig:exp}, another case of inter-class inconsistency in the performance of DU and UR50 is seen. There is an area of normal skin below the labeled skin lesion with a darker tone than the rest of the skin. This patch is incorrectly labeled as a skin lesion by DU and UR50. The UPCG and R2U models are the most robust in this experiment, while the change in skin tone also deceives R2U, although not as severely as DU and UR50. All in all, it is concluded that the proposed models that benefited from the attention modules can overcome some problems seen in the performance of other models. 

\begin{table}
\caption{Overall comparison of the models}
\centering
\label{table:overall}
\begin{tabular}{|l|ll|l|ll|}
\toprule
 {} & dice-score & iou-score & training-speed(\#epoch) & overfitting-dice & overfitting-iou \\
 \midrule
UR50 & \textbf{0.91±0.0} & \textbf{0.83±0.0} & 12.6 & \textbf{0.009±0.001} & \textbf{0.015±0.001} \\
R2U & 0.88±0.01 & 0.79±0.01 & \textbf{7.8} & 0.041±0.002 & 0.057±0.002 \\
R2UC & 0.88±0.01 & 0.79±0.01 & 8.8 & 0.048±0.007 & 0.066±0.01 \\
DU & 0.89±0.0 & 0.81±0.01 & 27.8 & 0.038±0.004 & 0.05±0.005 \\
\bottomrule
\end{tabular}
\end{table}

\section{CONCLUSION}
\label{sec:conclusion}
This work compares ten models with four augmentation setups to evaluate their performance and generalization using Dice Score, IoU, and a novel metric called $\Delta M $. Their training speed is also evaluated using a novel proposed method. Our experiments demonstrate the superiority of U-Net-Resnet50 and R2U-Net models with Aug-1 and Aug-2 augmentations regarding the metrics mentioned above and training speed, respectively. The Resnet50 backbone in the U-Net-Resnet50 is the main reason for its higher-level performance and generalization compared to other models. The well-trained backbone can extract essential information from the image to segment the skin lesions on the input images. The model R2U-Net, however, has the highest training speed, indicating that this model is suitable for applications with low-computational budgets.
Additionally, attention modules were intensively investigated in these models. Our experiments show that the CBAM module helped R2U-Net make more robust predictions, especially in the borders and areas with lighter color tones. The proposed U-Net Pyramid CBAM Gate scores lower according to the metrics mentioned before, but defeats U-Net-ResNet50, Double U-Net, and R2U-Net in some scenarios where inter-class indistinction and intra-class inconsistency can occur for these latter models. For future works, we will investigate different models using various backbones, especially Resnet and its variants. Furthermore, we will try to increase the performance of our proposed model with a diverse set of attention modules. Lastly, we would like to thank AIMedic Company\cite{AIMEDIC} for providing the required hardware and GPUs for training the models and for their helpful comments and insights on this work.

\bibliographystyle{elsarticle-num} 
\bibliography{main.bib}

\clearpage
\appendix

\section{more details}
\label{sec:appendix1}
\begin{figure}[ht]
\centering
\includegraphics[width=\linewidth]{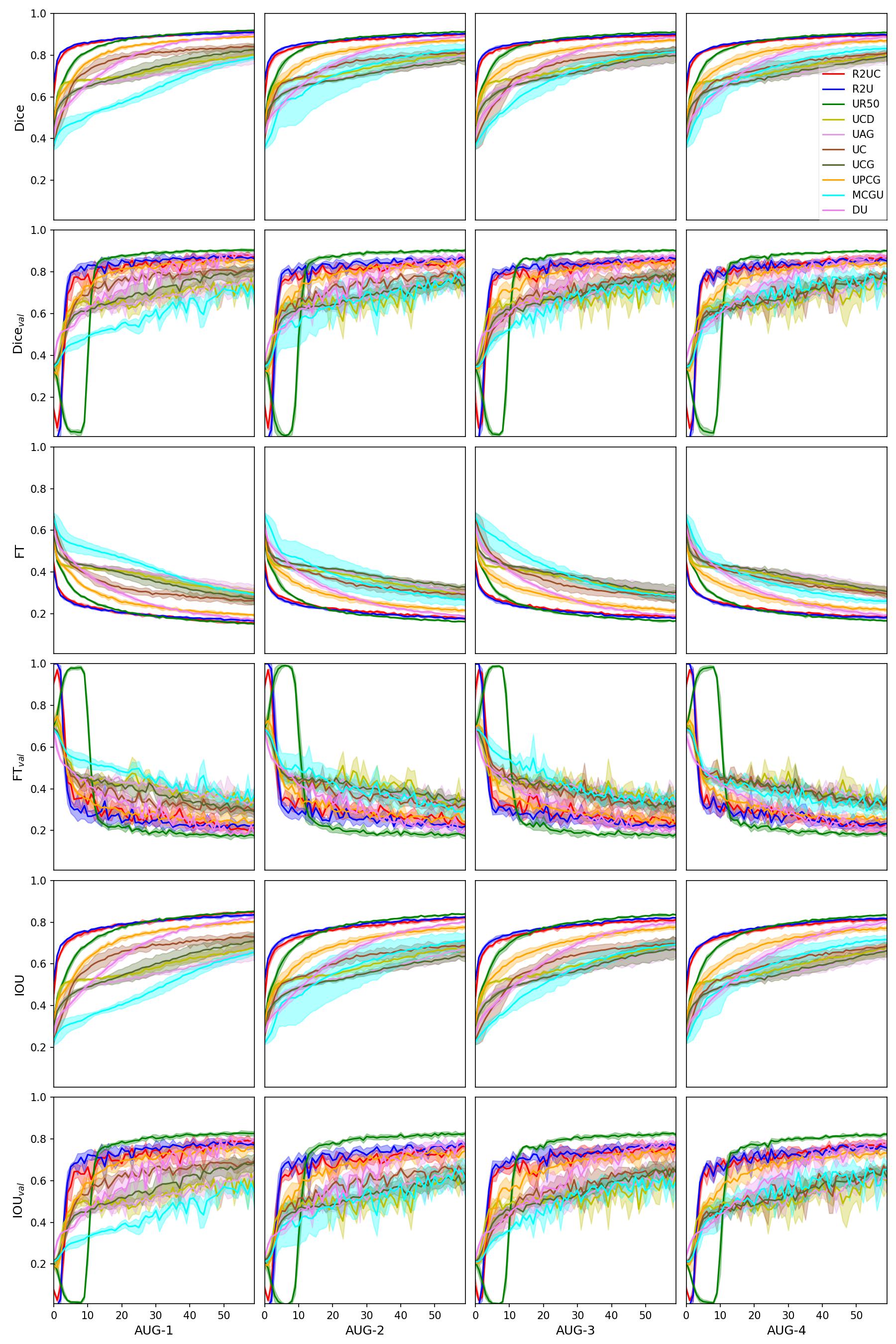}
\caption{}
\label{fig:all}
\end{figure}

\begin{figure}[ht]
\centering
\includegraphics[width=\linewidth]{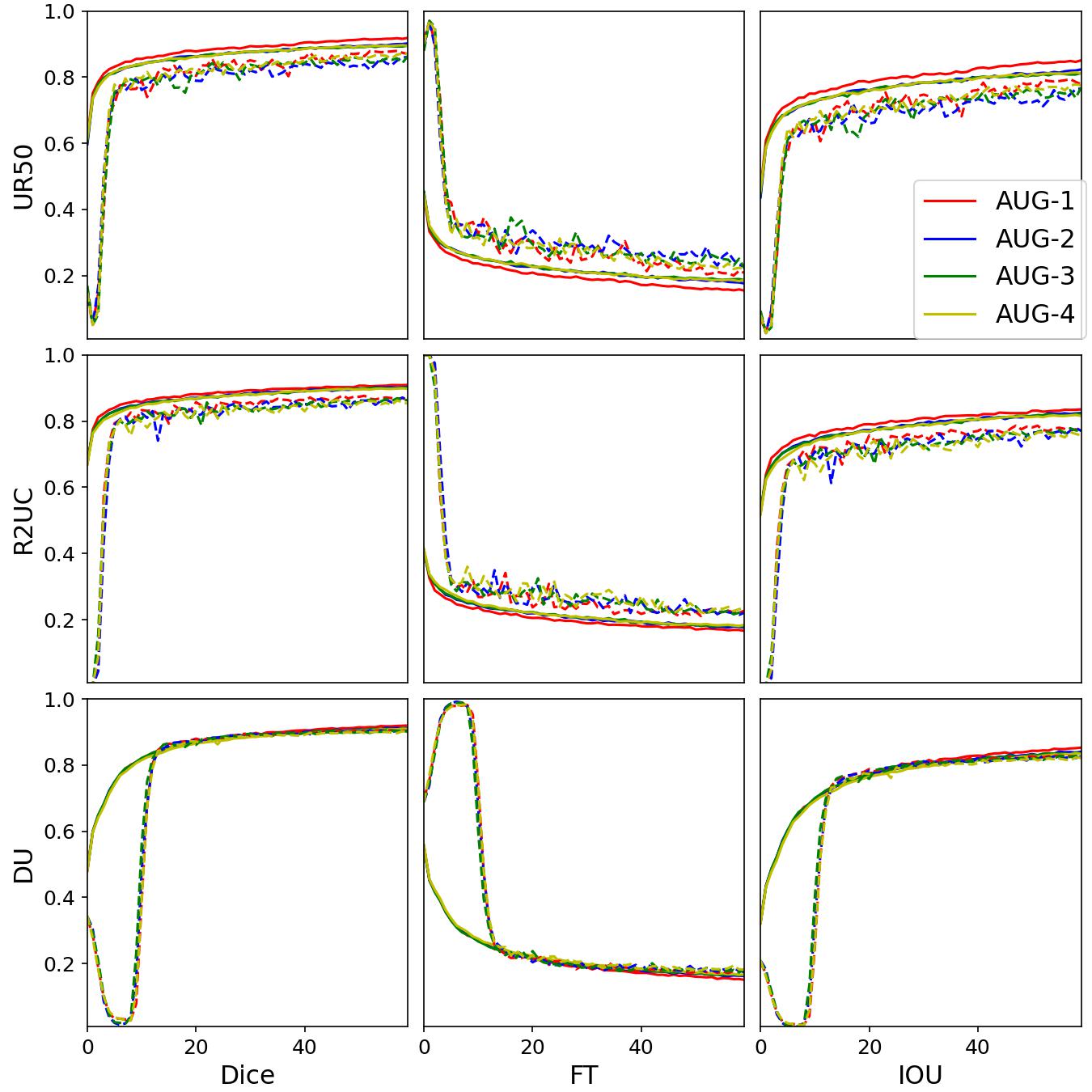}
\caption{Comparison of augmentation strategies through different loss functions and for the top-3 models. Solid and dashed lines show training and validation results, respectively.}
\label{fig:augs}
\end{figure}

In this section, all the results for ten models and four augmentations are presented in table~\ref{table:0}, table~\ref{table:1}, table~\ref{table:2}, and table~\ref{table:3}, respectively. In addition, the scores of all the models and four different augmentation configurations for different train and validation metrics are presented in figure~\ref{fig:all}. Finally, the same comparison is done for top-three models in figure~\ref{fig:augs}. 

\begin{table}
\centering
\caption{Models results for AUG-1}
\label{table:0}
\begin{tabular}{|l|ll|ll|ll|}
\toprule
{} &       dice &   val\_dice &         FT &     val\_FT &        iou &    val\_iou \\
\midrule
R2UC &   0.92±0.0 &  0.89±0.02 &   0.15±0.0 &  0.19±0.02 &   0.85±0.0 &   0.8±0.03 \\
R2U  &   0.91±0.0 &  0.89±0.01 &  0.16±0.01 &   0.2±0.02 &  0.84±0.01 &   0.8±0.02 \\
UR50 &   0.92±0.0 &   0.91±0.0 &   0.15±0.0 &  0.16±0.01 &   0.85±0.0 &  0.84±0.01 \\
U-Net  &   0.8±0.02 &  0.79±0.02 &   0.3±0.03 &  0.31±0.02 &  0.68±0.03 &  0.66±0.03 \\
UAG  &  0.79±0.03 &  0.78±0.05 &  0.31±0.03 &  0.31±0.05 &  0.66±0.04 &  0.65±0.07 \\
UC   &  0.85±0.02 &  0.83±0.02 &  0.26±0.02 &  0.28±0.02 &  0.74±0.02 &  0.71±0.02 \\
UCG  &  0.83±0.02 &  0.82±0.03 &  0.27±0.02 &  0.28±0.03 &  0.71±0.03 &  0.69±0.04 \\
UPCG &   0.89±0.0 &  0.86±0.02 &  0.19±0.01 &  0.22±0.03 &  0.81±0.01 &  0.77±0.02 \\
MCGU &  0.79±0.01 &  0.75±0.02 &  0.29±0.01 &  0.32±0.02 &  0.66±0.02 &  0.61±0.03 \\
DU   &    0.9±0.0 &   0.9±0.01 &  0.17±0.01 &  0.17±0.01 &  0.83±0.01 &  0.82±0.01 \\
\bottomrule
\end{tabular}
\end{table}

\begin{table}
\centering
\caption{Models results for AUG-2}
\label{table:1}
\begin{tabular}{|l|ll|ll|ll|}
\toprule
{} &       dice &   val\_dice &         FT &     val\_FT &        iou &    val\_iou \\
\midrule
R2UC &    0.9±0.0 &   0.87±0.0 &  0.17±0.01 &  0.21±0.01 &  0.83±0.01 &  0.78±0.01 \\
R2U  &   0.91±0.0 &  0.88±0.01 &   0.17±0.0 &   0.2±0.01 &   0.83±0.0 &  0.79±0.01 \\
UR50 &   0.91±0.0 &  0.91±0.01 &   0.16±0.0 &  0.16±0.01 &   0.84±0.0 &  0.83±0.01 \\
U-Net  &  0.81±0.02 &   0.8±0.02 &  0.29±0.02 &   0.3±0.03 &  0.68±0.03 &  0.67±0.03 \\
UAG  &   0.8±0.03 &  0.79±0.04 &   0.3±0.03 &  0.29±0.04 &  0.68±0.04 &  0.67±0.05 \\
UC   &  0.82±0.02 &  0.81±0.01 &  0.29±0.02 &  0.29±0.01 &  0.69±0.02 &  0.69±0.01 \\
UCG  &  0.78±0.02 &  0.77±0.03 &  0.32±0.02 &  0.32±0.02 &  0.64±0.03 &  0.63±0.04 \\
UPCG &  0.88±0.01 &  0.86±0.01 &  0.21±0.01 &  0.22±0.02 &  0.78±0.01 &  0.76±0.02 \\
MCGU &  0.83±0.03 &   0.8±0.03 &  0.26±0.03 &  0.28±0.04 &  0.72±0.05 &  0.67±0.04 \\
DU   &   0.89±0.0 &  0.89±0.01 &   0.19±0.0 &  0.18±0.01 &    0.8±0.0 &  0.81±0.01 \\
\bottomrule
\end{tabular}
\end{table}

\begin{table}
\centering
\caption{Models results for AUG-3}
\label{table:2}
\begin{tabular}{|l|ll|ll|ll|}
\toprule
{} &       dice &   val\_dice &         FT &     val\_FT &        iou &    val\_iou \\
\midrule
R2UC &    0.9±0.0 &  0.88±0.01 &  0.18±0.01 &  0.21±0.02 &  0.82±0.01 &  0.78±0.02 \\
R2U  &    0.9±0.0 &  0.87±0.01 &   0.18±0.0 &  0.21±0.02 &   0.82±0.0 &  0.78±0.01 \\
UR50 &   0.91±0.0 &   0.91±0.0 &   0.16±0.0 &  0.16±0.01 &  0.84±0.01 &  0.84±0.01 \\
U-Net  &  0.82±0.01 &   0.8±0.02 &  0.28±0.01 &   0.3±0.02 &   0.7±0.01 &  0.67±0.02 \\
UAG  &  0.81±0.03 &  0.79±0.04 &  0.29±0.03 &   0.3±0.05 &  0.68±0.04 &  0.66±0.06 \\
UC   &  0.82±0.03 &  0.81±0.02 &  0.28±0.04 &   0.3±0.03 &   0.7±0.04 &  0.68±0.03 \\
UCG  &   0.8±0.04 &   0.8±0.04 &  0.29±0.04 &   0.3±0.04 &  0.68±0.05 &  0.67±0.05 \\
UPCG &  0.88±0.01 &  0.86±0.01 &  0.21±0.01 &  0.22±0.02 &  0.78±0.01 &  0.76±0.02 \\
MCGU &  0.82±0.02 &  0.79±0.02 &  0.28±0.02 &   0.3±0.02 &   0.7±0.03 &  0.66±0.03 \\
DU   &   0.89±0.0 &  0.89±0.01 &  0.19±0.01 &  0.18±0.01 &   0.8±0.01 &  0.81±0.01 \\
\bottomrule
\end{tabular}
\end{table}

\begin{table}
\centering
\caption{Models results for AUG-4}
\label{table:3}
\begin{tabular}{|l|ll|ll|ll|}
\toprule
{} &       dice &   val\_dice &         FT &     val\_FT &        iou &    val\_iou \\
\midrule
R2UC &    0.9±0.0 &  0.88±0.01 &   0.18±0.0 &  0.21±0.02 &  0.82±0.01 &  0.79±0.02 \\
R2U  &    0.9±0.0 &   0.88±0.0 &   0.18±0.0 &   0.2±0.01 &   0.82±0.0 &  0.79±0.01 \\
UR50 &   0.91±0.0 &   0.91±0.0 &   0.16±0.0 &  0.17±0.01 &   0.84±0.0 &  0.83±0.01 \\
U-Net  &   0.8±0.02 &  0.79±0.03 &   0.3±0.02 &  0.31±0.03 &  0.68±0.03 &  0.66±0.04 \\
UAG  &   0.8±0.04 &  0.79±0.03 &   0.3±0.04 &  0.31±0.03 &  0.67±0.05 &  0.65±0.04 \\
UC   &  0.81±0.01 &  0.81±0.01 &  0.29±0.02 &   0.3±0.02 &  0.69±0.02 &  0.68±0.01 \\
UCG  &   0.8±0.03 &  0.79±0.02 &   0.3±0.03 &  0.31±0.02 &  0.67±0.04 &  0.66±0.03 \\
UPCG &  0.87±0.01 &  0.86±0.01 &  0.21±0.01 &  0.23±0.02 &  0.78±0.01 &  0.76±0.02 \\
MCGU &  0.84±0.02 &  0.81±0.04 &  0.25±0.02 &  0.28±0.04 &  0.72±0.03 &  0.69±0.05 \\
DU   &  0.89±0.01 &  0.89±0.01 &  0.19±0.01 &  0.18±0.01 &   0.8±0.01 &   0.8±0.01 \\
\bottomrule
\end{tabular}
\end{table}

% \end{thebibliography}
\end{document}